\documentclass{elsart}
\usepackage{amssymb,amsfonts}
\usepackage{graphicx}
\newcommand{\trace}{\mathop{\rm Tr}\nolimits}
\newcommand{\ddt}{\frac{d}{dt}\Big|_{t=0}}
\newcommand{\SD}{\mathop{\rm S}\nolimits}
\newcommand{\KL}{\mathop{\rm KL}\nolimits}
\newcommand{\JS}{\mathop{\rm JS}\nolimits}
\newcommand{\QJS}{\mathop{\rm QJS}\nolimits}

\newcommand{\diag}{\mathop{\rm Diag}\nolimits}

\newcommand{\supp}{\mathop{\rm supp}\nolimits}

\newcommand{\twomat}[4]{\left(\begin{array}{cc}#1&#2\\#3&#4\end{array}\right)}

\newcommand{\cD}{{\mathcal D}} % \cal not known
\newcommand{\cE}{{\mathcal E}} % \cal not known
\newcommand{\cT}{{\mathcal T}} % \cal not known
\newcommand{\cR}{{\mathcal R}} % \cal not known

\newcommand{\id}{\mathbb{I}}

\newcommand{\be}{\begin{equation}}
\newcommand{\ee}{\end{equation}}
\newcommand{\bea}{\begin{eqnarray}}
\newcommand{\eea}{\end{eqnarray}}
\newcommand{\beas}{\begin{eqnarray*}}
\newcommand{\eeas}{\end{eqnarray*}}
\newtheorem{definition}{Definition}
\newtheorem{theorem}{Theorem}
\newtheorem{lemma}{Lemma}

\newtheorem{proposition}{Proposition}

\newcount\minute
\newcount\hour
\def\currenttime{%
    \minute\time
    \hour\minute
    \divide\hour60
    \the\hour:\multiply\hour60\advance\minute-\hour\the\minute}
\begin{document}
\begin{frontmatter}
\title{Quantum Skew Divergence}
\author{Koenraad M.R.\ Audenaert}
\address{
Department of Mathematics,
Royal Holloway University of London, \\
Egham TW20 0EX, United Kingdom \\[1mm]
Department of Physics and Astronomy, University of Ghent, \\
S9, Krijgslaan 281, B-9000 Ghent, Belgium
}
\ead{koenraad.audenaert@rhul.ac.uk}
\date{\today, \currenttime}
%\begin{keyword}
%Matrix Inequality \sep Hamiltonian \sep Relative entropy \sep quantum information theory
%\MSC 15A60
%\end{keyword}
%------------------------------------------------------------------ ABSTRACT
\begin{abstract}
In this paper we study the quantum generalisation of the skew divergence, which is a dissimilarity measure between distributions
introduced by L.~Lee in the context of natural language processing.
We provide an in-depth study of the quantum skew divergence, including its relation to
other state distinguishability measures.
Finally, we present a number of important
applications: new continuity inequalities for the quantum Jensen-Shannon divergence and the Holevo information,
and a new and short proof of Bravyi's Small Incremental Mixing conjecture.
\end{abstract}

\end{frontmatter}
%%%%%%%%%%%%%%%%%%%%%%%%%%%%%%%%%%%%%%%%%%%%%%%%%%%%%%%%%%%%%%%%%%%%%%%%%%%%%%%%%%%%%%%%%%%%%
%%%%%%%%%%%%%%%%%%%%%%%%%%%%%%%%%%%%%%%%%%%%%%%%%%%%%%%%%%%%%%%%%%%%%%%%%%%%%%%%%%%%%%%%%%%%%
\section{Introduction\label{sec:intro}}
The quantum relative entropy of two density operators $\rho$ and $\sigma$, denoted $S(\rho||\sigma)=\trace\rho(\log\rho-\log\sigma)$,
was introduced by Umegaki \cite{umegaki} in 1962.
Since the 90's it gained in prominence, especially in the quantum information theory community,
when Hiai and Petz \cite{hiaipetz} showed that Umegaki's formula provided the proper quantum generalisation of the classical
Kullback-Leibler divergence $\KL(p||q)$ of two probability distributions, as an operational measure
of dissimilarity between quantum states. A lot of research has been spent exploring its mathematical and physical properties.
Despite having many universally useful features, the relative entropy exhibits certain properties that in some applications
may be considered as drawbacks. In particular, the relative entropy is not a distance measure in the mathematical sense of the word:
it is asymmetric with respect to interchanging arguments,
$S(\rho||\sigma)\neq S(\sigma||\rho)$, and it does not satisfy a triangle inequality. Moreover, the relative entropy is infinite whenever
the support of $\sigma$ is not contained in the support of $\rho$. This makes the relative entropy completely unsuitable as a distance
measure between pure states, for example. We will refer to this feature as the `infinity problem'.

Over the years, several modifications to the relative entropy have been proposed. Some of the better known modifications are the
Quantum Jensen-Shannon divergence \cite{roga12,fuglede04}, and the closely related Holevo information or
Holevo $\chi$ \cite{holevo,mikeandike}
(even though this is not usually considered as a
modification of the relative entropy in the QIT community because it serves entirely different purposes).

In the present paper we introduce another modification of the quantum relative entropy, which we call the
\textit{quantum skew divergence}.
We have coined this term\footnote{A preliminary version of this work has
already been presented at TQC-2011, Madrid \cite{QSD1},
but as we were then unaware of Lee's work the quantity came with another name, namely
`telescopic relative entropy'. We now feel that `quantum skew divergence' is a more informative name.}
because of its close similarity to the already existing classical concept of skew divergence of two probability
distributions, which was introduced by Lee \cite{lee99,lee01} in the context of natural language processing to overcome the infinity problem
for the Kullback-Leibler divergence.
As no confusion will arise we will henceforth refer to the quantum skew divergence as skew divergence (SD) for short.
It is not to be confused with the Wigner-Yanase-Dyson skew information and related notions, to which it bears no obvious resemblance.

The skew divergence is essentially the relative entropy but with `skewed' second argument. That is, the second argument $\sigma$ is replaced
by the convex combination $\alpha\rho+(1-\alpha)\sigma$, where $\alpha$ is a scalar ($0<\alpha<1$) which we call the \textit{skewing parameter}.
As one of its basic properties we will show that $S(\rho||\alpha\rho+(1-\alpha)\sigma)$ is no longer infinite but is bounded above by $-\log \alpha$,
and we define the skew divergence as the skewed relative entropy divided by this factor $-\log \alpha$:
$$
\SD_\alpha(\rho||\sigma) := \frac{1}{-\log \alpha} S(\rho||\alpha\rho+(1-\alpha)\sigma).
$$
Hence, $\SD_\alpha$ always takes values between 0 and 1.
It is to be noted that Lee's skew divergence does not have this normalisation factor.

This paper can be subdivided roughly in two parts: the first part is a theoretical study of the properties of the skew divergence,
and the second part is on applications.
The first part consists of six sections. After some preliminaries (Section \ref{sec:prelim}),
in Section \ref{sec:SD} we give precise definitions
for the skew divergence and state and prove its basic properties.

Sections \ref{sec:DSD} and \ref{sec:cont} are devoted to the more complicated continuity properties of the quantum skew divergence.
These are properties that have no counterparts for the relative entropy, as a direct consequence of the infinity problem.
First, we show that continuity holds in the sense that states that are close in trace norm distance are also close
when measured by the SD (Section \ref{sec:DSD}).
Secondly, we show that the SD is also continuous with respect to perturbations of each of its arguments (Section \ref{sec:cont}).
The proofs of these statements rely on some technical results
about the derivatives of the operator logarithm, and this is presented in Sections \ref{sec:log} and \ref{sec:proofs}.

In the second part of this paper we consider applications of the quantum skew divergence.
In Section \ref{sec:bravyi} we give a simple proof of the so-called Small Incremental Mixing Conjecture that was
postulated by Bravyi \cite{bravyi07} and recently proven by Van Acoleyen \cite{karel}. Our proof yields a better proportionality constant
(2 instead of 9) and may yield additional insight into the more general `mixing problem' proposed by Lieb and Vershynina \cite{lieb13}.

The second application (Section \ref{sec:dist}) is as a dissimilarity measure between quantum states,
being the original purpose for introducing the skew divergence. Here we give a detailed overview of the relative entropy's
drawbacks and of the various proposals that have been made in the literature and how the skew divergence fits in.

In Section \ref{sec:holevo} we note the close connection between SD and the generalised
quantum Jensen-Shannon divergence (QJS), i.e.\ the Holevo information. By exploiting the
sharp continuity estimates for the SD derived in this paper, we obtain new continuity-type bounds
for the QJS and the Holevo information that in many cases improve on existing estimates from the literature.

%%%%%%%%%%%%%%%%%%%%%%%%%%%%%%%%%%%%%%%%%%%%%%%%%%%%%%%%%%%%%%%%%%
\section{Preliminaries\label{sec:prelim}}
First, let us recall the definition of the quantum relative entropy \cite{ohya_petz,petzbook,wehrl}.
For quantum states $\rho$ and $\sigma$, both positive,
\be
S(\rho||\sigma) := \trace \rho(\log\rho-\log\sigma).
\ee
For non-normalised positive operators $A$ and $B$, one defines more generally
\be
S(A||B) := \trace A(\log A-\log B) - \trace(A-B).
\ee
For positive scalars $a$ and $b$, we will also write
\be
S(a|b) := a(\log a-\log b)-(a-b).
\ee
Strictly speaking, when $\sigma$ (or $B$) is not invertible, the quanutm relative entropy is no longer defined. However,
when the supports of $\rho$ and $\sigma$ satisfy the condition $\supp\rho\subseteq\supp\sigma$
one customarily adopts the convention that `$0^+\log 0^+ = 0$'
and redefines the relative entropy as
\beas
S(\rho||\sigma) &:=& S(\rho|_\sigma || \sigma|_\sigma),\\
S(A||B) &:=& S(A|_B || B|_B),
\eeas
where the symbol $A|_B$ denotes the restriction of $A$ to the support of $B$.
When $\supp\rho\not\subseteq\supp\sigma$ this redefinition is not possible and one says that the relative entropy is infinite,
leading to the infinity problem mentioned in the introduction.

Another important distance measure between density operators is the trace norm distance:
\[
T(\rho,\sigma):=\half ||\rho-\sigma||_1,
\]
where $||.||_1$ denotes the trace norm,
\[
||X||_1:= \trace |X| = \trace(X^*X)^{1/2}.
\]
For any self-adjoint operator $X$, let $X_+$ and $X_-$ denote the positive part $X_+ = (X+|X|)/2$
and negative part $X_- = (|X|-X)/2$; both parts are positive semidefinite\footnote{Note that the negative part is positive for the
same reason that the imaginary part of a complex number is real.}.
Then another expression for the trace norm distance is
$$
T(\rho,\sigma)= \trace(\rho-\sigma)_+ = \trace(\rho-\sigma)_-.
$$

%%%%%%%%%%%%%%%%%%%%%%%%%%%%%%%%%%%%%%%%%%%%%%%%%%%%%%%%%%%%%%%%%%%%%%%%%%%%%%%%%%%
\section{Quantum Skew Divergence\label{sec:SD}}
In this section we give a rigorous definition of the quantum generalisation of the skew divergence (SD) and state and prove its basic properties.

The quantum skew divergence is based on the functional $S(\rho||\alpha\rho+(1-\alpha)\sigma)$,
or $S(A||\alpha A+(1-\alpha)B)$ in the non-normalised case,
where $\alpha$ is a scalar, with $0<\alpha<1$.
Since, for all such $\alpha$, $\supp(A)\subseteq\supp(A+B)=\supp(\alpha A+(1-\alpha)B)$, no problem of infinities arises.
Henceforth, we will always write $S(A||\alpha A+(1-\alpha)B)$, whether $A,B>0$ or $A,B\ge0$. In the latter case this is
to mean $S(A|_{A+B} || (\alpha A+(1-\alpha)B)|_{A+B})$.
\begin{definition}
For fixed $\alpha\in(0,1)$, the \emph{quantum $\alpha$-skew divergence} between states $\rho$ and $\sigma$ is defined as
\be
\SD_\alpha(\rho||\sigma) := \frac{1}{-\log(\alpha)} \,\,S(\rho||\alpha\rho+(1-\alpha)\sigma).
\ee
Likewise, for non-normalised operators $A,B\ge0$,
\be
\SD_\alpha(A||B) := \frac{1}{-\log(\alpha)} \,\,S(A||\alpha A+(1-\alpha)B).
\ee
\end{definition}
We call $\alpha$ the \textit{skewing parameter}.

The reason for incorporating the scale factor $1/(-\log \alpha)$ is to normalise the range of the SD to the interval $[0,1]$.
\begin{theorem}\label{th:01}
For all states $\rho$ and $\sigma$ and $0<\alpha<1$,
$$
0\le \SD_\alpha(\rho||\sigma)\le 1,
$$
and
$\SD_\alpha(\rho||\sigma)= 1$ if and only if $\rho\perp\sigma$.
\end{theorem}
Recall that two quantum states are mutually orthogonal, denoted $\rho\perp\sigma$, iff $\trace\rho\sigma=0$.

\noindent\textit{Proof.}
Let $\tau=\alpha\rho+(1-\alpha)\sigma$.
By operator monotonicity of the logarithm, we have
$$
\log(\tau)=\log(\alpha\rho+(1-\alpha)\sigma)\ge \log(\alpha\rho),
$$
and, therefore,
\beas
S(\rho||\tau)&=&\trace\rho(\log\rho-\log\tau) \\
&\le& \trace\rho(\log\rho-\log(\alpha\rho)) \\
&=& -\log \alpha.
\eeas
Thus, $S(\rho||\tau)$ is bounded above by $-\log \alpha$, which is finite for $0<\alpha<1$.
It therefore makes perfect sense to normalise $S(\rho||\tau)$ by dividing it by $-\log \alpha$, producing a quantity
that is always between $0$ and $1$.

The equality case was proven in \cite{QSD1}.
\qed

The definition of the skew divergence for non-normalised operators is also applicable to non-negative scalars. To distinguish
the scalar case more clearly from the matrix case we will use the symbol $\SD_\alpha(b|c)$ for scalars; we have
\be
\SD_\alpha(b|c) = \frac{b(\log b - \log(\alpha b+(1-\alpha)c)) - (1-\alpha)(b-c)}{-\log \alpha}.
\ee

As we do not restrict the arguments of the SD to be normalised states, the following \textit{scaling identities}
can be useful.
\begin{theorem}
For $0<\alpha<1$, operators $X,Y\ge0$, and positive scalars $b,c$,
\bea
\SD_\alpha(bX||bY) &=& b \SD_\alpha(X||Y)\\
\SD_\alpha(bX||cX) &=& \SD_\alpha(b|c)\trace X.
\eea
\end{theorem}
This is easy to prove by simple calculation.

The quantum skew divergence inherits many desirable properties from the quantum relative entropy:
\begin{theorem}
For $0<\alpha<1$, states $\rho$, $\sigma$, any unitary matrix $U$ and any completely positive trace-preserving (CPTP) map $\Phi$,
\begin{enumerate}
\item \emph{Positivity:} $\SD_\alpha(\rho||\sigma)\ge0$, and $\SD_\alpha(\rho||\sigma)=0$ if and only if $\rho=\sigma$;
\item \emph{Unitary invariance:} $\SD_\alpha(U\rho U^*||U\sigma U^*)=\SD_\alpha(\rho||\sigma)$;
\item \emph{Contractivity:} $\SD_\alpha(\Phi(\rho)||\Phi(\sigma))\le \SD_\alpha(\rho||\sigma)$;
\item \emph{Joint convexity:} the map $(\rho,\sigma)\mapsto\SD_\alpha(\rho||\sigma)$ is jointly convex.
\end{enumerate}
\end{theorem}
The proof is again straightforward.
Note that these are the same properties that
the quantum Jensen-Shannon divergence obeys \cite{lamberti05}.

\section{The Operator Logarithm and its Derivatives\label{sec:log}}
%The projector on the support of $X$ is denoted by $\{X\}$.
The following integral representation of the logarithm lies at the basis of much of the subsequent
treatment. For $x>0$, we have
\be
\log x = \int_0^\infty ds \left(\frac{1}{1+s}-\frac{1}{x+s}\right).\label{eq:intlog}
\ee
Using functional calculus, this definition can be extended to the operator logarithm. For $A>0$,
\be
\log A = \int_0^\infty ds \left(\frac{1}{1+s}\id-(A+s\id)^{-1}\right).\label{eq:intlogop}
\ee
From this representation follow representations
of the first and second derivatives of the operator logarithm.
\subsection{First Derivative}
Following \cite{lieb73}, let us define for $A>0$ the linear map $\Delta\to\cT_A(\Delta)$ for self-adjoint $\Delta$ as
the Fr\'echet derivative of the operator logarithm:
\be
\cT_A(\Delta) := \frac{d}{dt}\Bigg|_{t=0} \log(A+t\Delta).\label{eq:cTdef}
\ee
From integral representation (\ref{eq:intlog})
we get an integral representation for $\cT_A$ as well:
\be
\cT_A(\Delta) = \int_0^\infty ds\,\,(A+s\id)^{-1} \Delta (A+s\id)^{-1}.\label{eq:cTintrep}
\ee
Here we have used the fact that
$$
\frac{d}{dt} (A+t\Delta)^{-1} = -(A+t\Delta)^{-1}\Delta(A+t\Delta)^{-1}.
$$
Being a positive linear combination of conjugations it follows that, for any  $A>0$, $\cT_A$ is a completely positive map. In particular, it preserves the
positive semidefinite order; that is, if $X\le Y$, then $\cT_A(X)\le\cT_A(Y)$.
Also, $X>0$ implies $\cT_A(X)>0$.

\begin{lemma}
For $A>0$ and $\Delta=\Delta^*$, and scalars $a>0$ and $\delta$,
\be
\cT_{aA}(\delta\Delta) = \frac{\delta}{a}\cT_A(\Delta).
\ee
Furthermore,
\be
\cT_A(A)=\id.
\ee
\end{lemma}
\textit{Proof.}
For the first identity:
\beas
\cT_{aA}(\delta\Delta) &=& \ddt\log(aA+t\delta\Delta) = \ddt\log(A+t(\delta/a)\Delta) \\
&=& \cT_{A}((\delta/a)\Delta) = (\delta/a)\cT_{A}(\Delta).
\eeas
The second identity follows similarly from the fact that $\log(A+tA)=(1+t)\id+\log A$, and the term $\log A$ drops out after differentiating.
\qed

Hence,
for scalar arguments we have
\be
\cT_a(\delta) = \delta/a.
\ee

\begin{lemma}\label{lem:Tab}
For $A,B\ge0$ with $A+B>0$
\be
\cT_{A+B}(A) \le \id.
\ee
\end{lemma}
\textit{Proof.}
Since $B\ge 0$, we have
$A+B \ge A$
and because $\cT_{A+B}$ preserves the positive semidefinite ordering,
$\cT_{A+B}(A) \le \cT_{A+B}(A+B) = \id$.
\qed
%%%%%%%%%%%%%%%%%%%%%%%%%%%%%%%%%
\subsection{The metric $M_A(B,C)$}
The sesquilinear form
\be
M_A(B,C):=\langle B^*,\cT_A(C)\rangle=\trace B^*\cT_A(C)
\ee
which is defined for $A>0$,
is a \textit{metric}: it is self-adjoint ($M_A(B,C) = \overline{M_A(C,B)}$),
positive semidefinite ($M_A(B,B)\ge0$ for any $B$),
with $M_A(B,B)=0$ iff $B=0$, and $M_A(B,B)$ is continuous in $B$ for any $A$.
As the metric is contractive under completely positive trace-preserving (CPTP) maps $\Phi$,
$$
M_{\Phi(A)}(\Phi(B),\Phi(B))\le M_A(B,B)
$$
for any $A>0$ and any $B$, it is a \textit{monotone metric} \cite{lesniewski,petz96}.
%This particular sesquilinear form is closely related to the \textit{Kubo-Mori scalar product} from quantum statistical mechanics,
%which is the sequilinear form $\langle B^*,\cT^{-1}_A(C)\rangle$.
Lieb has shown that the map $(A,B)\mapsto M_A(B,B)$, for $A>0$ and any $B$,
is jointly convex in $A$ and $B$ (\cite{lieb73}, Theorem 3).
%Further properties are discussed in \cite{lieb73}.

$M$ satisfies the following limit property:
\begin{lemma}\label{lem:epsilon}
Let $A,B,C\ge0$ with $B+C>0$ and $\supp A\subseteq\supp B$.
Then%, irrespective of $\supp C$,
$$
\lim_{\epsilon\to 0} M_{B+\epsilon C}(A,A) = M_{B|_B}(A|_B,A|_B).
$$
\end{lemma}
\textit{Proof.}
%To begin with, we can restrict $A$ and $B$ in the left-hand side to $\supp(B+C)$.
Let $P$ be the projector on $\supp B$ and $Q$ the
projector on the orthogonal complement of $\supp B$. % (within $\supp (B+C)$).
Consider the $2\times2$ partitioning induced by $P$ and $Q$:
$$
A\to\twomat{PAP^*}{PAQ^*}{QAP^*}{QAQ^*},
$$
and similarly for all other operators.
Because of the conditions on the supports, we have $PAQ^*=QAP^*=QAQ^*=0$ and $PBQ^*=QBP^*=QBQ^*=0$.
Hence,
\beas
\lefteqn{\trace A\cT_{B+\epsilon C}(A)} \\
&=& \int_0^\infty ds\;\trace A(B+\epsilon C+s)^{-1}A(B+\epsilon C+s)^{-1} \\
&=& \int_0^\infty ds\;\trace (PAP^*)\;(P(B+\epsilon C+s)^{-1}P^*)\;(PAP^*)\;(P(B+\epsilon C+s)^{-1}P^*).
\eeas
Using Schur complements, we can find the explicit expression
\beas
\lefteqn{P(B+\epsilon C+s)^{-1}P^*} \\
&=& \left((PBP^* + \epsilon PCP^*+s) - \epsilon^2 PCQ^*(\epsilon QCQ^*+s)^{-1}QCP^*\right)^{-1}.
\eeas
In the limit $\epsilon\to0$, this simplifies as
$$
\lim_{\epsilon\to0}P(B+\epsilon C+s)^{-1}P^* = (PBP^*+s)^{-1},
$$
since all operator blocks appearing here are invertible.
Therefore,
\beas
\lim_{\epsilon\to0}\trace A\cT_{B+\epsilon C}(A)
&=& \int_0^\infty ds\;\trace (PAP^*)\;(PBP^*+s)^{-1}\;(PAP^*)\;(PBP^*+s)^{-1} \\
&=& \trace A|_B\cT_{B|_B}(A|_B).
\eeas
\qed
%%%%%%%%%%%%%%%%%%%%%%%%%%%%%%%%%
\subsection{Second Derivative}
Having defined the linear operator $\cT$ via the first derivative of the logarithm, we can also define
a quadratic operator $\cR$ via the second derivative \cite{lieb73}. For $A>0$ and $\Delta$ self-adjoint, let
\be
\cR_A(\Delta) := -\frac{d^2}{dt^2}\Bigg|_{t=0}\log(A+t\Delta).\label{eq:defRA}
\ee
A simple calculation using the integral representation of the first derivative yields the
integral representation
\be
\cR_A(\Delta) = 2\int_0^\infty ds\,\,(A+s\id)^{-1}\Delta(A+s\id)^{-1}\Delta(A+s\id)^{-1}.
\label{eq:intcR}
\ee
One can similarly define a bilinear form, for $A>0$ and self-adjoint $\Delta_1$ and $\Delta_2$:
\bea
\cR_A(\Delta_1,\Delta_2) &:=&  -\frac{d^2}{dt_1dt_2}\Bigg|_{t_1=t_2=0}\log(A+t_1\Delta_1+t_2\Delta_2) \\
&=& \int_0^\infty ds\,\,(A+s\id)^{-1}\Delta_1(A+s\id)^{-1}\Delta_2(A+s\id)^{-1} \nonumber\\
&& \mbox{}+\int_0^\infty ds\,\,(A+s\id)^{-1}\Delta_2(A+s\id)^{-1}\Delta_1(A+s\id)^{-1}.
\eea
Clearly,
\bea
\cR_A(\Delta,\Delta) &=& \cR_A(\Delta),\\
\cR_A(\Delta_1,\Delta_2) &=& \cR_A(\Delta_2,\Delta_1),\\
\trace \Delta_0\cR_A(\Delta_1,\Delta_2) &=& \trace \Delta_2\cR_A(\Delta_0,\Delta_1).
\eea

It is readily checked that for scalar $a$ and $\delta$ we have
\be
\cR_{aA}(\delta\Delta) = (\delta/a)^2 \cR_A(\Delta)
\ee
and
\be
\cR_a(\delta) = (\delta/a)^2.
\ee

\begin{lemma}\label{lem:RATA}
For $A>0$ and $\Delta=\Delta^*$
\[
\cR_A(A,\Delta)=\cT_A(\Delta).
\]
Hence
\[
\cR_A(A)=\id.
\]
\end{lemma}
\textit{Proof.}
\beas
\cR_A(A,\Delta)
&=& -\frac{d^2}{dt_1 dt_2}\Bigg|_{t_1=t_2=0} \log(A+t_1A+t_2\Delta) \\
&=& -\frac{d^2}{dt_1 dt_2}\Bigg|_{t_1=t_2=0} \log(1+t_1)\id+\log(A+t_2/(1+t_1)\Delta)\\
&=& -\frac{d}{dt_1}\Bigg|_{t_1=0} \frac{d}{dt_2}\Bigg|_{t_2=0} \log(A+t_2/(1+t_1)\Delta).
\eeas
The derivative w.r.t.\ $t_2$ is, with $u=t_2/(1+t_1)$,
\beas
\frac{d}{dt_2}\Bigg|_{t_2=0} \log(A+t_2/(1+t_1)\Delta)
&=& \frac{d}{du}\Bigg|_{u=0} \log(A+u\Delta)\frac{1}{1+t_1} \\
&=& \cT_A(\Delta) \frac{1}{1+t_1}.
\eeas
Therefore,
$$
\cR_A(A,\Delta)
= -\frac{d}{dt_1}\Bigg|_{t_1=0} \cT_A(\Delta) \frac{1}{1+t_1} = \cT_A(\Delta).
$$
\qed

\begin{lemma}\label{lem:lieb2}
For $A,B\ge0$, with $A+B>0$,
\be
\cR_{A+B}(A)\le\id.
\ee
\end{lemma}
\textit{Proof.}
Due to the bilinearity of $\cR_A(\Delta_1,\Delta_2)$ and Lemma \ref{lem:RATA}, we have
\beas
\cR_{A+B}(A) &=& \cR_{A+B}(A+B-B) = \cR_{A+B}(A+B-B,A+B-B) \\
&=& \cR_{A+B}(A+B,A+B) + \cR_{A+B}(-B,A+B) \\
&& \mbox{}+ \cR_{A+B}(A+B,-B) + \cR_{A+B}(-B,-B) \\
&=& \id - 2\cT_{A+B}(B) + \cR_{A+B}(B).
\eeas
The third term can be bounded in terms of the second.
Since $A+B+s\id\ge B$, for any $s\ge0$, we have $(A+B+s\id)^{-1}\le B^{-1}$ and $B(A+B+s\id)^{-1}B\le B$.
Therefore,
\beas
\cR_{A+B}(B) &=& 2\int_0^\infty ds\,\,(A+B+s\id)^{-1}\,\,B(A+B+s\id)^{-1}B\,\,(A+B+s\id)^{-1} \\
&\le&2\int_0^\infty ds\,\,(A+B+s\id)^{-1}\,\,B\,\,(A+B+s\id)^{-1} \\
&=& 2\cT_{A+B}(B).
\eeas
We finally get
\[
\cR_{A+B}(A) \le \id-2\cT_{A+B}(B)+2\cT_{A+B}(B) = \id.
\]
\qed

%%%%%%%%%%%%%%%%%%%%%%%%%%%%%%%%%%%%%%%%%%%%%%%%%%%%%%%%%%%%%%%%%
%%%%%%%%%%%%%%%%%%%%%%%%%%%%%%%%%%%%%%%%%%%%%%%%%%%%%%%%%%%%%%%%%
\section{A Continuity Inequality for the metric $M$\label{sec:proofs}}
In this section we will prove the following technical inequality for the metric $M$, which will be used heavily in the proofs of the continuity
properties of the quantum skew divergence.
\begin{theorem}\label{th:Tderiv}
For $A,B,C\ge0$ with $A+B>0$, and with $a=\trace A$, $c=\trace C$,
\be
0 \le M_{A+B}(A,A) - M_{A+B+C}(A,A) \le M_{a}(a,a) - M_{a+c}(a,a)
\ee
or, explicitly,
\be
0 \le \trace A \cT_{A+B}(A) - \trace A \cT_{A+B+C} (A) \le a - \frac{a^2}{a+c}.\label{eq:Tderiva}
\ee
\end{theorem}

To prove this theorem, we need the following lemma:
\begin{lemma}\label{lem:f}
Let $f(t)$ be a real-valued convex function on $[0,1]$.
If, moreover, $f(0)\le0$ and $f(0)\le f'(0)$,
then $\forall t\in[0,1], f(0)\le (1-t)f(t)$.
\end{lemma}
\textit{Proof.}
Since $f(0)\le0$, for all $t\in[0,1]$ we have $f(0) \le f(0)(1-t) \le f'(0)(1-t)$.
Multiplying both sides by $t$ and adding $(1-t)f(0)$ gives
$f(0) \le t(1-t)f'(0) +(1-t)f(0) = (1-t)(f(0)+tf'(0))$.
By convexity of $f$, $f(0)+tf'(0)$ is a lower bound on $f(t)$,
and the inequality of the lemma follows.
\qed

\textit{Proof of Theorem \ref{th:Tderiv}.}
The first inequality in (\ref{eq:Tderiva}) easily follows from the fact that $x\mapsto 1/x$ is
operator monotone decreasing together with the identity
$$
\trace X \cT_A(X) = \int_0^\infty ds \,\,\trace(X^{1/2}(A+s\id)^{-1} X^{1/2})^2,
$$
and monotonicity of the function $X\to \trace X^2$.

%%%%%%%%%%%%%%%%%%%%%%%%%%%%%%%%
The second inequality involves more work.
Let us thereto consider two positive density operators $\rho$ and $\sigma$, an operator $G\ge\rho$,
and the function
\beas
f(t)&=& -1+\frac{d}{ds}\Bigg|_{s=0} \trace \rho \log(G+s\rho +t(\sigma-G))\\
&=& \trace\rho\cT_{t\sigma + (1-t)G}(\rho)-1.
\eeas
We start by showing that $(1-t)f(t)\ge f(0)$ for $0\le t\le 1$.

\noindent Firstly, $f(0)=\trace\rho\cT_G(\rho)-1$. Since $\cT_G(\rho) \le \cT_G(G)=\id$, we have $f(0)\le0$.

\noindent Secondly, the derivative $f'(0)$ is given by
\beas
f'(0) &=& \frac{d^2}{ds dt}\Bigg|_{s=t=0}  \trace \rho \log(G+s\rho +t(\sigma-G)) \\
&=& \trace \rho \cR_G(\rho,G-\sigma) = \trace \rho \cR_G(\rho,G)-\trace \rho \cR_G(\rho,\sigma).
\eeas
The first term can be rewritten as
\[
\trace \rho\cR_G(G,\rho) = \trace \rho\cT_G(\rho),
\]
by Lemma \ref{lem:RATA}.
Because $G\ge\rho$, the second term can be bounded using Lemma \ref{lem:lieb2} as
\[
\trace \rho \cR_G(\rho,\sigma) = \trace\sigma \cR_G(\rho) \le\trace\sigma = 1.
\]
We therefore obtain
\[
\trace \rho \cR_G(\rho,G-\sigma) \ge \trace \rho\cT_G(\rho)-1,
\]
which proves that $f'(0)\ge f(0)$.

By Lieb's convexity theorem, the map $G\mapsto \trace\rho\cT_G(\rho)$ is convex, hence $f(t)$ is convex.

All three conditions of Lemma \ref{lem:f} are therefore satisfied, so that $(1-t)f(t)\ge f(0)$, for $0\le t\le 1$.

Now let $a>0$, $c\ge0$, and $G=\rho+B/a$,
with $B\ge 0$; this choice indeed satisfies the condition $G\ge\rho$.
With this substitution, we get
\[
f(t) = \trace \rho \cT_{(1-t)\rho+\frac{1}{a}(1-t)B+t\sigma}(\rho) - 1.
\]
In particular, with the choice $t=c/(a+c)$,
\beas
(1-t)f(t)
&=& \frac{a}{a+c} \left(\trace \rho \cT_{\frac{a}{a+c}\rho+\frac{1}{a+c}B+\frac{c}{a+c}\sigma}(\rho) - 1\right) \\
&=& \trace \rho \cT_{\rho+\frac{1}{a}B+\frac{c}{a}\sigma}(\rho) - \frac{a}{a+c} \\
f(0) &=& \trace \rho \cT_{\rho+\frac{1}{a}B}(\rho) - 1.
\eeas
The inequality $(1-t)f(t)\ge f(0)$ therefore gives (after multiplying by $a$)
Multiplying by $a$ yields
\[
\trace a\rho \cT_{a\rho+B+c\sigma}(a\rho) - \frac{a^2}{a+c} \ge \trace a\rho \cT_{a\rho+B}(a\rho) - a.
\]
or, after rearranging terms,
$$
\trace a\rho\cT_{a\rho+B}(a\rho) - \trace a\rho\cT_{a\rho+B+c\sigma}(a\rho) \le a - \frac{a^2}{a+c}.
$$
Setting $A=a\rho$ and $C=c\sigma$ we obtain the second inequality of (\ref{eq:Tderiva}).
\qed
%%%%%%%%%%%%%%%%%%%%%%%%%%%%%%%%%%%%%%%%%%%%%%%%%%%%%%%%%%%%%%%%%%%%%%%%%%%%%%%%%%%%%%%%%%%%%%%%%%%%%%%%%%%%%%
\section{Quantum Skew Divergence as Integral of the Metric $M$\label{sec:DSD}}
The reason for considering the metric $M$ in such detail as we have done, is that the quantum skew divergence can be conveniently
written as an integral of $M$. To this purpose, let us introduce the following quantity based on $M$, which
can be seen as a differential version of the SD:
\begin{definition}
Let $A,B\ge0$ such that $A+B>0$.
For $0<\alpha<1$, define
\be
\cD_\alpha(A||B) := \alpha(1-\alpha) M_{\alpha A+(1-\alpha)B}(A-B,A-B).\label{eq:DSDM}\label{eq:DSDexpl2}
\ee
For $\alpha=0,1$, define $\cD_\alpha$ to be identically zero.

For general $A,B\ge0$ not satisfying the condition $A+B>0$, define
$\cD_\alpha(A||B)$ as $\cD_\alpha(A|_{A+B}||B|_{A+B})$.
\end{definition}
Other explicit formulas for $\cD_\alpha$ are:
\bea
\cD_\alpha(A||B) &=& \alpha(\trace A \cT_{\alpha A+(1-\alpha)B}(A-B) - \trace(A-B)) \label{eq:DSDexpl1} \\
&=& \frac{\alpha}{1-\alpha} \trace A\cT_{\alpha A+(1-\alpha)B}(A) - \frac{\alpha}{1-\alpha}\trace A -\alpha\trace(A-B).\label{eq:DSDexpl3}
\eea
These formulas follow from (\ref{eq:DSDM}) by expressing $A-B$ as
\[
A-B = \frac{1}{1-\alpha}(A-(\alpha A+(1-\alpha)B))
\]
and exploiting the identities $\cT_X(X)=\id$ and $\trace A \cT_X(B) = \trace B \cT_X(A)$.

We denote $\cD_\alpha$ for scalar arguments by $\cD_\alpha(b|c)$. Explicit formulas are
\bea
\cD_\alpha(b|c) &=& \alpha(1-\alpha)\frac{(b-c)^2}{\alpha b+(1-\alpha)c} \label{eq:DSDscalar} \\
&=& \frac{\alpha}{1-\alpha}\left(\frac{b^2}{\alpha b+(1-\alpha)c} - b\right)-\alpha(b-c).\label{eq:DSDscalar2}
\eea
In particular,
\be
\cD_\alpha(b|0) = (1-\alpha)b,\qquad \cD_\alpha(0|c)=\alpha c.
\ee

From the properties of $M$, it follows that $\cD_\alpha$ is positive and contractive under CPTP maps.
For example, with $a=\trace A$ and $b=\trace B$, we have:
\be
\cD_\alpha(A||B) \le \cD_\alpha(a|b).
\ee

Clearly, $\cD_\alpha$ is unitarily invariant:
for any unitary $U$, $\cD_\alpha(UAU^*||UBU^*) = \cD_\alpha(A||B)$.

A very useful property of $\cD_\alpha$ is the following \textit{symmetry property}.
\begin{theorem}\label{th:DSDsymm}
For $A,B\ge0$, and $0< \alpha< 1$,
\be
\cD_\alpha(A||B) = \cD_{1-\alpha}(B||A).
\ee
\end{theorem}
\textit{Proof.}
This follows immediately from formula (\ref{eq:DSDexpl2}).
\qed

\bigskip

%----------------------------------------------------- Connection to Kullback-Leibler
We will now show how the quantum skew divergence is related to $\cD_\alpha$.
It is well-known that the quantum relative entropy $S(A||B)$ is differentiable w.r.t.\ $A$ and $B$ whenever $A,B>0$.
Hence, for $A,B>0$, the function $\alpha\mapsto S(A||\alpha A+(1-\alpha)B)$ is differentiable over the open interval $(0,1)$.
For $A,B\ge0$ this is no longer true as the relative entropy is in general only lower semicontinuous \cite{wehrl}.
However, if one restricts $A$ and $B$ to the support of $A+B$, % in the more general case $A,B\ge0$,
the function $\alpha\mapsto S(A||\alpha A+(1-\alpha)B)$ is still differentiable for $A,B\ge0$.
Because of this, the following connection between $\cD_\alpha$ and $\SD_\alpha$ emerges:
\begin{lemma}
For $A,B\ge0$ and $0<\alpha<1$,
\bea
\cD_\alpha(A||B) &=& \frac{d}{d(-\log \alpha)}S(A||\alpha A+(1-\alpha)B) \\
&=& -\alpha\;\frac{d}{d\alpha}S(A||\alpha A+(1-\alpha)B).
\eea
\end{lemma}

Conversely, $\SD_\alpha$ can be obtained from $\cD_\alpha$ by a simple \textit{averaging procedure}.
\begin{theorem}\label{th:average}
For operators $A,B\ge0$ and $0<\alpha<1$,
\be
\SD_\alpha(A||B) = \frac{1}{-\log \alpha} \int_0^{-\log \alpha} \cD_{\alpha'}(A||B) \; d(-\log\alpha').\label{eq:average}
\ee
\end{theorem}
\textit{Proof.}
Define the function $f(\alpha) = S(A||\alpha A+(1-\alpha)B)$. By the substitution $b=-\log \alpha$, we can write
\beas
\SD_\alpha(A||B) &=& \frac{1}{b} f(\exp(-b)) \\
\cD_\alpha(A||B) &=& \frac{d}{db} f(\exp(-b)).
\eeas
Therefore, as for $b=0$, $f(\exp(-b))=f(1)=S(A||A)=0$,
\beas
\SD_\alpha(A||B) &=& \frac{1}{b} \int_0^b \frac{d}{db} f(\exp(-b))\;db \\
&=& \frac{1}{-\log \alpha} \int_0^{-\log \alpha} \cD_{\alpha'}(A||B) \; d(-\log\alpha'),
\eeas
which is indeed an average w.r.t.\ $-\log \alpha$.
\qed

This is an important fact, because whenever one has an equality or inequality involving several instances of $\cD_\alpha$
with the same value of $\alpha$,
one can immediately obtain the corresponding (in)equality for $\SD_\alpha$ by averaging over a suitable
range of $-\log \alpha$.

%-----------------------------------------------------

To end this section, we use the averaging technique to derive sharp inequalities relating $\SD_\alpha(\rho,\sigma)$ to the trace norm distance
$T(\rho,\sigma)$.
We will encounter further applications of this technique in the proofs of Proposition \ref{prop:rbts2} and Theorems \ref{th:triangle1} and \ref{th:bravyi}.

The quantity $\cD_\alpha$ is related to one of the so-called
\textit{quantum $\chi^2$-divergences} introduced by Temme \textit{et al} \cite{temme}, namely the one induced by the
logarithm. This logarithmic quantum $\chi^2$-divergence is defined for $A,B>0$ as
$$
\chi^2_{\log}(A,B) := M_B(A-B,A-B) = \trace(A-B)\cT_B(A-B).
$$
A short calculation reveals that
\be
\cD_\alpha(A||B) = \frac{\alpha}{1-\alpha}\chi^2_{\log} (A, \alpha A+(1-\alpha)B).
\ee
This means that certain properties that were proven in \cite{temme} for the quantum $\chi^2$-divergences carry over to $\cD_\alpha$.
One such property is the following lower bound on $\cD_\alpha$ in terms of the trace norm distance $T(\rho,\sigma)$:
\begin{theorem}
For all density operators $\rho$ and $\sigma$ and any $0<\alpha<1$,
\be
\cD_\alpha(\rho||\sigma) \ge 4\alpha(1-\alpha) T(\rho,\sigma)^2.
\ee
\end{theorem}
\textit{Proof.}
This follows from Lemma 5 in \cite{temme} according to which $\chi^2(\rho,\sigma)\ge ||\rho-\sigma||_1^2$.
With the substitution $\sigma\to \tau:=\alpha\rho+(1-\alpha)\sigma$ and noting that $\rho-\tau = (1-\alpha)(\rho-\sigma)$, the inequality follows.
\qed

We can also furnish an upper bound on $\cD_\alpha$ in terms of
the trace norm distance.
\begin{theorem}
For density operators $\rho,\sigma\ge0$ and $0< \alpha< 1$,
\be
\cD_\alpha(\rho||\sigma) \le T(\rho,\sigma).
\ee
\end{theorem}
\textit{Proof.}
%For $0<a<1$,
From formula (\ref{eq:DSDexpl1}) and the basic properties of $\cT$,
\beas
\cD_\alpha(\rho||\sigma)
&=& \alpha\trace \rho \cT_{\alpha\rho+(1-\alpha)\sigma}(\rho-\sigma) \\
&=& \alpha\trace (\rho-\sigma) \cT_{\alpha\rho+(1-\alpha)\sigma}(\rho) \\
&\le& \alpha\trace (\rho-\sigma)_+ \cT_{\alpha\rho+(1-\alpha)\sigma}(\rho) \\
&\le& \trace (\rho-\sigma)_+ \cT_{\alpha\rho+(1-\alpha)\sigma}(\alpha\rho+(1-\alpha)\sigma) \\
&=& \trace(\rho-\sigma)_+ = T(\rho,\sigma).
\eeas
\qed

Using the averaging procedure, Theorem \ref{th:average}, we immediately get the promised relations for $\SD_\alpha$:
\begin{theorem}\label{th:SDvT}
For density operators $\rho,\sigma\ge0$ and $0< \alpha< 1$,
\be
\frac{2(1-\alpha)^2}{-\log(\alpha)} T(\rho,\sigma)^2\le \SD_\alpha(\rho||\sigma) \le T(\rho,\sigma).
\ee
\end{theorem}
To prove the lower bound we note that using (\ref{eq:average})
the factor $4\alpha(1-\alpha)$ averages to $2(1-\alpha)^2/(-\log(\alpha))$.

The upper bound shows that two states that are close in trace norm distance are also close
in terms of $\SD_\alpha$.
Despite the very simple form of the upper bound, it is the strongest one possible.
Equality can be obtained for any value of $t=T(\rho,\sigma)$ for states in dimension 3 (and higher),
for example by choosing
$\rho=\diag(t,0,1-t)$ and $\sigma=\diag(0,t,1-t)$.

%------------------------------------------------------------------------------------
\section{Continuity Properties of the Quantum Skew Divergence\label{sec:cont}}
The inequalities of Theorem \ref{th:Tderiv} lead to several inequalities for $\cD_\alpha$, which in turn lead to inequalities
for the quantum skew divergence.
\begin{theorem}\label{th:DSDdiff}
For $A,B,C\ge0$ and $0< \alpha< 1$, with $a=\trace A$ and $c=\trace C$,
\bea
-\alpha c = -\cD_\alpha(0|c) &\le& \cD_\alpha(A||B) - \cD_\alpha(A||B+C) \nonumber\\
&\le& \cD_\alpha(a|0) - \cD_\alpha(a|c).\label{eq:DSDdiff2}
\eea
\bea
0 &\le& \cD_\alpha(B||A+B) - \cD_\alpha(B+C||A+B+C) \nonumber \\
&\le& \cD_\alpha(0|a) - \cD_\alpha(c|a+c).\label{eq:rbts2DSD}
\eea
\end{theorem}
\textit{Proof.}
Consider first the case $A,B,C>0$ of inequalities (\ref{eq:DSDdiff2}).
These follow from Theorem \ref{th:Tderiv} and expressions (\ref{eq:DSDexpl3}) and (\ref{eq:DSDscalar2}).
We have
\beas
\lefteqn{\cD_\alpha(A||B) - \cD_\alpha(A||B+C)} \\
&=& \frac{\alpha}{1-\alpha}\left(M_{\alpha A+(1-\alpha)B}(A,A) - M_{\alpha A+(1-\alpha)(B+C)}(A,A)\right) - \alpha\trace C \\
&=& \frac{1}{1-\alpha}\left(M_{A+\frac{1-\alpha}{\alpha}B}(A,A) - M_{A+\frac{1-\alpha}{\alpha}B+\frac{1-\alpha}{\alpha}C}(A,A)\right) - \alpha\trace C.
\eeas
The first term is now of the form that allows Theorem \ref{th:Tderiv} to be invoked and (\ref{eq:DSDdiff2}) follows immediately.

To treat the case $A,B,C\ge0$ we use
Lemma \ref{lem:epsilon} to bring both terms on a `common denominator' as far as supports are concerned.
Whereas $\cD_\alpha(A||B)$ is defined as $\cD_\alpha(A|_{A+B}||B|_{A+B})$, and in the second term the operators are
restricted to the potentially larger subspace $\supp(A+B+C)$, we can write
$$
\cD_\alpha(A||B) - \cD_\alpha(A||B+C) = \lim_{\epsilon\to0} \cD_\alpha(A||B+\epsilon C) - \cD_\alpha(A||B+C),
$$
in which the operators in both terms are now restricted to the support of $A+B+C$,
allowing to use the positive case, as before.

To prove the second set of inequalities (\ref{eq:rbts2DSD}) we
use the expression (\ref{eq:DSDM}) and the substitution $A'=(1-\alpha)A$ (so $\trace A'=(1-\alpha)\trace A$):
\beas
\lefteqn{\cD_\alpha(B||A+B) - \cD_\alpha(B+C||A+B+C)} \\
&=& \alpha(1-\alpha)\left(M_{\alpha B+(1-\alpha)(A+B)}(A,A) - M_{\alpha (B+C)+(1-\alpha)(A+B+C)}(A,A)\right) \\
&=& \frac{\alpha}{1-\alpha}\left(M_{A'+B}(A',A') - M_{A'+B+C}(A',A')\right),
\eeas
which is again of the form required by Theorem \ref{th:Tderiv}.
\qed

Equality in the lower bounds of (\ref{eq:DSDdiff2}) and (\ref{eq:rbts2DSD}) is attained for $A=\diag(a,0)$, $B=\diag(b_1,b_2)$ and $C=\diag(0,c)$.
Equality in the upper bounds is attained for scalar $A,B,C$. Thus, the given bounds are the best possible among all bounds that are only based on
$a,c$ and $\alpha$.

Theorem \ref{th:DSDdiff} immediately yields:
\begin{proposition}\label{prop:rbts2}
For operators $A,B,C\ge0$, with $a=\trace A$ and $c=\trace C$,
\bea
-\SD_\alpha(0|c)&\le& \SD_\alpha(A||A+B)-\SD_\alpha(A||A+B+C) \le -\SD_\alpha(a|a+c) \label{eq:rbtsSD}\\
-S(0|c)&\le& S(A||A+B)-S(A||A+B+C) \le -S(a|a+c).\label{eq:rbts}
\eea
\bea
0&\le &\SD_\alpha(B||A+B) - \SD_\alpha(B+C||A+B+C) \nonumber \\
 &\le& \SD_\alpha(0|a)-\SD_\alpha(c|a+c) \label{eq:rbts2SD}\\
0&\le &S(B||A+B) - S(B+C||A+B+C) \nonumber \\
&\le& S(0|a)-S(c|a+c).\label{eq:rbts2}
\eea
\end{proposition}
\textit{Proof.}
Inequalities (\ref{eq:rbtsSD}) follow by averaging those of (\ref{eq:DSDdiff2}) and noting that $\SD_\alpha(a|a)=0$.
Inequalities (\ref{eq:rbts2SD}) follow by averaging those of (\ref{eq:rbts2DSD}).

Then note that
\beas
(-\log\alpha)\SD_\alpha(A||A+B)
&=& S(A||\alpha A +(1-\alpha)(A+B)) \\
&=& S(A||A+(1-\alpha)B).
\eeas
Doing this for all the terms in (\ref{eq:rbtsSD}) and absorbing the factors $(1-\alpha)$ in $B$, $C$ and $c$ yields (\ref{eq:rbts}).
A similar procedure yields (\ref{eq:rbts2}) from (\ref{eq:rbts2SD}).
\qed

From Theorem \ref{th:DSDdiff}, it is easy to derive quantitative continuity properties for $S_\alpha$. The following theorem gives bounds on
the change of $\SD_\alpha$ (and $\cD_\alpha$) when either of its arguments changes, as expressed by the trace distance.
Here we restrict to density operators (trace equal to 1).
\begin{theorem}\label{th:triangle1}
Let $0< \alpha< 1$.

For density operators $\rho,\sigma_1,\sigma_2$ such that $T(\sigma_1,\sigma_2)=t$,
\bea
|\cD_\alpha(\rho||\sigma_1) - \cD_\alpha(\rho||\sigma_2)| &\le& \cD_\alpha(1|0) - \cD_\alpha(1 | t) +\cD_\alpha(0|t)\label{eq:tri_int}\\
|\cD_\alpha(\sigma_1||\rho) - \cD_\alpha(\sigma_2||\rho)| &\le& \cD_\alpha(0|1) - \cD_\alpha(t | 1) +\cD_\alpha(t|0)\label{eq:tri_int2}
\eea
and
\bea
|\SD_\alpha(\rho||\sigma_1) - \SD_\alpha(\rho||\sigma_2)| &\le& \SD_\alpha(1|0) - \SD_\alpha(1|t) +\SD_\alpha(0|t)\label{eq:triangle1}\\
|\SD_\alpha(\sigma_1||\rho) - \SD_\alpha(\sigma_2||\rho)| &\le& \SD_\alpha(0|1) - \SD_\alpha(t | 1) +\SD_\alpha(t|0).\label{eq:triangle2}
\eea
\end{theorem}
\textit{Proof.}
Let $A,B_1,B_2\ge0$.
A successive application of the first and then the second inequality of (\ref{eq:DSDdiff2}) yields
\beas
\lefteqn{\cD_\alpha(A||B_1) - \cD_\alpha(A||B_2)} \\
&=& \cD_\alpha(A||B_1) - \cD_\alpha(A||B_1 + (B_2-B_1)_+ - (B_2-B_1)_-) \\
&\le& \cD_\alpha(A||B_1) - \cD_\alpha(A||B_1 + (B_2-B_1)_+) +\cD_\alpha(0|\trace(B_2-B_1)_-)\\
&\le& \cD_\alpha(\trace A|0) - \cD_\alpha(\trace A | \trace(B_2-B_1)_+) +\cD_\alpha(0|\trace(B_2-B_1)_-).
\eeas
Specialising to $A=\rho$ and $B_i=\sigma_i$, with
$\trace(\sigma_2-\sigma_1)_+=\trace(\sigma_2-\sigma_1)_- =:t$, we get (\ref{eq:tri_int}).
Inequality (\ref{eq:tri_int2}) follows immediately from (\ref{eq:tri_int}) by the symmetry of $\cD_\alpha$ (Theorem \ref{th:DSDsymm}).
Using the averaging procedure we get the same inequalities with $\cD_\alpha$ replaced by $\SD_\alpha$, giving (\ref{eq:triangle1})
and (\ref{eq:triangle2}).
Due to the symmetry under exchanging $\sigma_1$ and $\sigma_2$ we can add an absolute value sign to the left-hand side of all these inequalities.
\qed

\textbf{Remarks.}
\begin{enumerate}
\item It can be checked that the right-hand side of inequality (\ref{eq:triangle1}) is a concave and monotonously increasing function
of $t$ for any $0<\alpha<1$.
\item It is also easily verified that equality is achieved in (\ref{eq:triangle1}) for $\rho\perp\sigma_1$ and $\sigma_2=t\rho+(1-t)\sigma_1$.
\item Unlike in Proposition \ref{prop:rbts2}, this approach does not lead to corresponding inequalities for the relative entropy proper, $S$,
as no such inequalities can exist.
Indeed, no matter how small $t$ is,
one can always find states $\rho$, $\sigma_1$ and $\sigma_2$ such that $|S(\rho||\sigma_1)-S(\rho||\sigma_2)|$ is unbounded; take, for example,
$\rho=\sigma_2$ and $\sigma_1$ such that $\supp\rho$ is not a subspace of $\supp\sigma_1$.
\end{enumerate}
%%%%%%%%%%%%%%%%%%%%%%%%%%%%%%%%%%%%%%%%%%%%%%%%%%%%%%%%%%%%%%%%%%%%%%%%%%%%%%%%%%%%%%%%%%%%%%%%%%
%%%%%%%%%%%%%%%%%%%%%%%%%%%%%%%%%%%%%%%%%%%%%%%%%%%%%%%%%%%%%%%%%%%%%%%%%%%%%%%%%%%%%%%%%%%%%%%%%%
\section{The Small Incremental Mixing Conjecture\label{sec:bravyi}}
Consider an ensemble of time-dependent states, $\cE(t)=\{(p_j, \rho_j(t))\}_{j=1}^n$,
where each state $\rho_j(t)$ evolves under the influence of a Hamiltonian $H_j$; that
is, $\rho_j(t) = U_j(t)\rho_j U_j(t)^*$, where $U_j(t)=\exp(itH_j)$.
Let $\rho_0(t)$ be the ensemble averaged state, $\rho_0(t) = \sum_{j=1}^n p_j\rho_j(t)$.
We will drop the time argument to indicate the state at time 0, $\rho_j := \rho_j(0)$.

The \textit{mixing rate} $\Lambda(\cE)$ of this ensemble is defined as
$$
\Lambda(\cE) := \frac{d}{dt}\Bigg|_{t=0}S(\rho_0(t)).
$$
Bravyi conjectured in \cite{bravyi07} the following upper bound on the mixing rate for binary ensembles ($n=2$):
$$
\Lambda(\cE) \le c\; h_2(p)\; ||H_1-H_2||,
$$
where $c$ is a dimension- and state-independent constant, and $h_2(p)$ is the Shannon entropy of the distribution $(p,1-p)$.
He called this the \textit{Small Incremental Mixing} (SIM) conjecture.
Lieb and Vershynina considered this conjecture in \cite{lieb13} and
inquired whether this bound could also be valid for larger ensembles ($n>2$); that is, whether
$$
\Lambda(\cE) \le c\; H(\mathbf{p}),
$$
where $H(\mathbf{p})$ is the Shannon entropy of the ensemble's probability vector, and all the Hamiltonians satisfy
$||H_j||\le 1$.

Bravyi's SIM conjecture was proven very recently by Van Acoleyen \textit{et al} \cite{karel}, with a value for the constant $c=9$.
More details about the physical relevance of this conjecture (now a theorem), in particular
to entanglement generating rates and entanglement area laws, can be found in \cite{bravyi07,lieb13,karel}.

In this Section we provide an entirely different proof, and obtain a sharper form of the inequality, with constant $c=2$.
Our approach is based on the observation that the mixing rate can be expressed in terms of $\SD_\alpha$.
Without loss of generality we can put $H_1=0$ and replace $H_2$ by $H$, so that $U_1(t)=\id$, $U_2(t)=U(t)$ and $\rho_1(t)=\rho_1$.
Because the entropy of
the signal states $\rho_j(t)$ does not change under unitary evolution,
we have
\bea
\lefteqn{S(\rho_0(t)) - S(\rho_0)} \nonumber\\
&=&\left(S(\rho_0(t)) - \sum_j p_j S(\rho_j(t))\right)-\left(S(\rho_0) - \sum_j p_j S(\rho_j)\right)\nonumber\\
&=& \sum_j p_j \left(S(\rho_j(t)||\rho_0(t)) - S(\rho_j||\rho_0)\right) \nonumber\\
&=& -p_1 \log(p_1) (\SD_{p_1}(\rho_1||\rho_2(t)) - \SD_{p_1}(\rho_1||\rho_2)) \nonumber\\
&&  -p_2 \log(p_2) (\SD_{p_2}(\rho_2(t)||\rho_1) - \SD_{p_2}(\rho_2||\rho_1)) \nonumber\\
&=& -p_1 \log(p_1) (\SD_{p_1}(\rho_1||U(t)\rho_2 U^*(t)) - \SD_{p_1}(\rho_1||\rho_2)) \nonumber\\
&&  -p_2 \log(p_2) (\SD_{p_2}(\rho_2||U^*(t)\rho_1 U(t)) - \SD_{p_2}(\rho_2||\rho_1)).\label{eq:SvSD}
\eea
In the last line we have exploited unitary invariance of $\SD_\alpha$.

A natural first attempt is to try inequality (\ref{eq:triangle1}) of Theorem \ref{th:triangle1} (with $s=0$).
\beas
S(\rho_0(t)) - S(\rho_0)&\le& -\sum_j p_j\log(p_j) (\SD_{p_j}(1|0) - \SD_{p_j}(1|t_j) + \SD_{p_j}(0|t_j)) \\
&\le& -\sum_j p_j\log(p_j) \frac{1-p_j}{-p_j\log(p_j)}t_j,
\eeas
where $t_1 = T(U(t)\rho_2 U^*(t), \rho_2)$ and
$t_2 = T(U^*(t)\rho_1 U(t), \rho_1)$.
This requires estimating the trace norm distances $t_j$
but it can already be seen that we will obtain a bound that is too weak, due to the
occurrence of the factor $(1-p_j)/(-p_j\log(p_j))$, which can become arbitrarily large for small $p_j$.

The following theorem is a substantial sharpening of inequality (\ref{eq:triangle1})
for the special case that $\sigma_1$ and $\sigma_2$ are unitarily equivalent.
\begin{theorem}\label{th:bravyi}
For states $\rho$ and $\sigma$, for $0<\alpha<1$, and $U=\exp(iH)$,
\be
\SD_\alpha(\rho||U\sigma U^*) - \SD_\alpha(\rho||\sigma) \le 2||H||.\label{eq:SDbravyi}
\ee
\end{theorem}
This is the key result leading to our proof of the SIM conjecture.

The proof of this theorem relies on the following simple estimate of the
trace norm distance between two unitarily equivalent states.
\begin{lemma}\label{lem:evol}
For a state $\rho$ subject to a unitary evolution $U(t)=\exp(itH)$,
\be
T(U(t)\rho U^*(t),\rho) \le t\,||H||.
\ee
\end{lemma}
\textit{Proof.}
Let $\rho'=U(t)\rho U^*(t)$.
For infinitesimal $dt$,
$U=\id+i \,dt\, H$ and $U\rho U^*=\rho+i\,dt\,[H,\rho]$.
Thus $||\rho'-\rho||_1 = dt\,\,||\,[H,\rho]\,||_1\le dt\,\,2||H||\,\, ||\rho||_1$,
where we used the triangle inequality for the trace norm, and H\"older's inequality.
Integrating over $t$ and using the triangle inequality once more
shows that this is also true for finite $t$.
\qed

\textit{Proof of Theorem \ref{th:bravyi}.}
Rather than working with $\SD_\alpha$, we consider $\cD_\alpha$ because its symmetry property is essential.
For all density operators $\rho$, $\sigma_1$ and $\sigma_2$, and $0<\alpha<1$, with $\tau=T(\sigma_1,\sigma_2)$, inequality
(\ref{eq:tri_int}) reads
\beas
\cD_\alpha(\rho||\sigma_1) - \cD_\alpha(\rho||\sigma_2) &\le& \cD_\alpha(1|0) -\cD_\alpha(1|\tau)+\cD_\alpha(0|\tau) \\
&=& \frac{\tau}{\alpha+(1-\alpha)\tau} \le \frac{\tau}{\alpha}.
\eeas
In particular, for $\sigma_2=\sigma$ and $\sigma_1 = U\sigma U^*$, with $U=\exp(iH)$,
$$
\cD_\alpha(\rho||U\sigma U^*) - \cD_\alpha(\rho||\sigma) \le \frac{1}{a} T(U\sigma U^*,\sigma) \le \frac{1}{\alpha} ||H||,
$$
where we also have used Lemma \ref{lem:evol}.

From the symmetry property of $\cD_\alpha$, Theorem \ref{th:DSDsymm},
it follows that the inequality also holds when replacing $\alpha$ in the right-hand side by $1-\alpha$.
Indeed,
\beas
\cD_\alpha(\rho||U\sigma U^*) - \cD_\alpha(\rho||\sigma)
&=& \cD_{1-\alpha}(U\sigma U^*||\rho) - \cD_{1-\alpha}(\sigma||\rho) \\
&=& \cD_{1-\alpha}(\sigma||U^*\rho U) - \cD_{1-\alpha}(\sigma||\rho) \\
&\le& \frac{1}{1-\alpha} T(U^*\rho U,\rho) \le \frac{1}{1-\alpha}||H||.
\eeas
Hence, combining the two inequalities yields
$$
\cD_\alpha(\rho||U\sigma U^*) - \cD_\alpha(\rho||\sigma) \le \min\left( \frac{1}{\alpha},\frac{1}{1-\alpha}\right) ||H|| \le 2||H||.
$$
Using the averaging procedure then yields the inequality of the theorem.
\qed

\begin{theorem}[Small Incremental Mixing]\label{th:SIMg}
Within the setup described above,
\be
S(\rho_0(t)) - S(\rho_0) \le 2t \;h(p_1,p_2)  ||H||.
\ee
\end{theorem}
\textit{Proof.}
To each term of (\ref{eq:SvSD}) we apply Theorem \ref{th:bravyi} to estimate the differences between the $\SD_\alpha$ and get
\beas
S(\rho_0(t)) - S(\rho_0) &\le& -\sum_{j=1}^2 p_j\log(p_j) \;2t||H|| = 2t\; h(p_1,p_2) \; ||H||.
\eeas
\qed

%%%%%%%%%%%%%%%%%%%%%%%%%%%%%%%%%%%%%%%%%%%%%%%%%%%%%%%%%%%%%%%%%%%%%%%%%%%%%%%%%%%%%%%%%%%%%%%%%%
%%%%%%%%%%%%%%%%%%%%%%%%%%%%%%%%%%%%%%%%%%%%%%%%%%%%%%%%%%%%%%%%%%%%%%%%%%%%%%%%%%%%%%%%%%%%%%%%%%%%%%%%%%%%%%%%%%%
\section{Quantum Skew Divergence as a State Distinguishability Measure\label{sec:dist}}
The quantum relative entropy (QRE) between two quantum states $\rho$ and $\sigma$,
$S(\rho||\sigma)=\trace\rho(\log\rho-\log\sigma)$, is a non-commutative generalisation
of the Kullback-Leibler divergence (KLD) $\KL(p||q)$ between probability distributions $p$ and $q$, and
is widely used as a measure of dissimilarity of
quantum states \cite{ohya_petz}.

Both the KLD and the QRE exhibit a number of features that arise naturally from their underlying mathematical model and
that may be useful in certain circumstances. However, these features also
imply that neither the KLD nor the QRE is a proper distance measure in the mathematical sense.
First of all, the KLD and QRE are asymmetric in their arguments. This alone already precludes their use as a distance measure, and
prompted the terminology KL `divergence', rather than KL `distance'.
Secondly, neither obeys the triangle inequality.
A third feature, and the one considered in this paper,
is that the KLD is infinite whenever for some $i$, the probability $q(i)$ is zero when $p(i)$ is not.
Likewise, $S(\rho||\sigma)$ is infinite when the support of $\rho$ is not contained in the support of $\sigma$.
In particular, this renders the relative entropy useless as a useful distance measure between pure states,
since it is infinite for pure $\rho$ and $\sigma$, unless
$\rho$ and $\sigma$ are exactly equal (in which case it always gives $0$). It is therefore unable to tell by how much
two distinct pure states are dissimilar.

It is illustrative to see how this feature comes about in one of the more important operational interpretations of the KLD and QRE,
namely in the context of asymmetric hypothesis testing.
Let the null hypothesis $H_0$ be that a random variable $X$ is drawn from the distribution $p$; the alternative hypothesis $H_1$,
that it is drawn from distribution $q$. A test is to be designed that optimally discriminates between the two.
Two types of error are relevant: a type I error (false positive) is when the test selects $H_1$ when in fact $H_0$ is true;
a type II error (false negative) is when the test selects $H_0$ when $H_1$ is true.
The probability of a type I error is usually denoted by $\alpha$, and the probability of a type II error by $\beta$.
These probabilities cannot usually both be made zero, but they can be made to both tend to 0 exponentially fast
when $N$, the number of samples of $X$ looked at by the test, tends to infinity.
One can then define the corresponding error rates, $\alpha_R$ and $\beta_R$, as the limits
$\alpha_R = -\lim_{N\to \infty}(1/N)\log\alpha_N$ and
$\beta_R = -\lim_{N\to \infty}(1/N)\log\beta_N$. These rates quantify how fast $\alpha_N$ and $\beta_N$ tend to 0 with $N$.

The KLD can be given a clear operational meaning in this context, as the best possible rate $\beta_R$ when $\alpha_N$ (not $\alpha_R$)
is to be kept below a certain value $\epsilon$ (a value which, surprisingly, does not ultimately
enter in the value of the optimal $\beta_R$).
It is now not hard to see why the KLD should be infinite when, for some $i$, $q(i)$ is zero but $p(i)$ is not.
In this case an optimal test should only look at outcome $i$. If this outcome occurs, even if only once, this immediately
rules out the alternative hypothesis. The number of samples required to find outcome $i$ amongst them
(which depends on $p(i)$) is finite, therefore
the rate $\beta_R$ is infinite. In other words, the infinity of the KLD represents the fact
that ``the theory `All crows are black' can be refuted by the single observation of a
white crow''.

Whereas the emergence of this feature of the KLD (and the QRE) seems quite natural, it may not always be that desirable.
Firstly, the unboundedness of the KLD may be a source of numerical instability in applications.
Secondly, the extreme focus on zeros of $q$ (zero eigenvalues of $\sigma$, respectively) implies a complete disregard
of other discriminating information. As stated before, the QRE can only tell distinctness of pure states, but not by how much.
Thirdly, in applications where $q$ is an \emph{empirical} distribution, the weight put on events with $q(i)=0$ is totally inappropriate:
in empirical distributions this corresponds to unseen events, not to impossible ones. This is a serious concern in applications such as
natural language processing \cite{lee99}, where the events are occurrences of word combinations in a large (but not infinitely large)
corpus of text, and in which many genuine but rare word combinations do not occur at all\footnote{Consider, for example,
the total number of occurrences
of the word combination ``relative entropy'' in the combined issues of the New York Times.}.
Similar concerns can be raised
in the quantum case, when $\sigma$ is a reconstructed quantum state obtained from quantum state tomography experiments.
When maximum likelihood reconstruction of nearly pure states produces reconstructed states with one or more zero eigenvalues,
these zeroes should not be interpreted as zero probabilities. How to properly deal with these empirical quantum states is known
in the tomography literature as the `zero-eigenvalue problem' \cite{robin}.
A final problem is of a theoretical nature: because KLD and QRE can become infinite, it is much harder (and less natural) to
obtain good upper bounds on these quantities in terms of other distance measures. Invariably, some information about the smallest
eigenvalues of $\rho$ and $\sigma$ have to be supplied to allow even the existence of such bounds (see, e.g.\ \cite{ka2,ka1}).

Several solutions have been put forward to overcome the problems associated with this infinity feature, in the classical case
and in the quantum case, in the form of modifications of the KLD (QRE).
In the classical case, one of the first to discuss several of these modifications in detail was Lin \cite{lin91}.
In addition to the infinity problem, he also considered the asymmetry issue.
He introduced the following dissimilarity measures based on the KLD, which he called the \textit{$K$-divergence}
and \textit{$L$-divergence}, respectively:
\bea
K(p||q) &=& S(p||(p+q)/2) \\
L(p,q) &=& K(p||q)+K(q||p) \\
&=& 2H((p+q)/2)-H(p)-H(q).
\eea
Here, $H(p)$ is the Shannon entropy of a distribution, which for the discrete case reads $H(p)=-\sum_i p(i)\log p(i)$.
Lin also considered a generalisation of the $L$-divergence with different weights, which he called the
\textit{Jensen-Shannon divergence}:
\be
\JS^\alpha(p,q) = H(\alpha p+(1-\alpha)q)-\alpha H(p)-(1-\alpha)H(q).
\ee
Lin pointed out that the $K$ divergence is a special case of the Csisz\'ar $f$-divergences with the function $f$ given by
$f(x)=x\log(2x/(1+x))$ \cite{csiszar}.

In \cite{lee99}, Lee introduced a generalisation of Lin's $K$-divergence that incorporates different weights,
\be
s_\alpha(p||q) = S(p||\alpha q+(1-\alpha)p),
\ee
a quantity which she called the \textit{$\alpha$-skew divergence}.
In contrast to Lin's, whose motivations were mainly theoretical and driven by the lack of good upper bounds on the KL divergence,
Lee's proposal was fuelled by a practical application in natural language processing: the estimation and subsequent use
of probabilities of unseen word combinations \cite{lee99,lee01}.
Here, the asymmetry of the KLD had to be maintained but
its inordinate focus on zero-probabilities had to be alleviated. Lee proposed a `smoothing' of the $q$ distribution with $p$ by mixing
a small amount of $p$ into $q$ (she used $\alpha=0.99$), in order to shift the focus to events that are seen under both distributions.

In the quantum case, the first attempt to overcome the infinity problem of the QRE was undertaken
by Lendi, Farhadmotamed and van~Wonderen
\cite{lendi}, who proposed to mix both $\rho$ and $\sigma$ with the maximally mixed state. They introduced
the \textit{regularised relative entropy} as
$$
R(\rho||\sigma)=c_d\,\, S\left(\frac{\rho+\id_d}{1+d} \Bigg|\Bigg|\frac{\sigma+\id_d}{1+d}\right),
$$
where $d$ is the dimension of state space, and $c_d$ is a normalisation constant.
It is clear that this procedure only works for finite-dimensional states.
One might also consider mixing both states with a smaller amount of the maximally mixed state, for example as a
quantum generalisation of Laplace's rule of succession for empirical distributions, by which 1 is added to the frequencies
of all outcomes, in order to properly account for unseen events.

Another possibility, also applicable to the infinite dimensional case,
is to apply a smoothing process. One can define the \textit{smooth relative entropy}
between states $\rho$ and $\sigma$ as
the infimum of the ordinary relative entropy between $\rho$ and another (unnormalised) state $\tau$, where $\tau$
is constrained to be $\epsilon$-close to $\sigma$ in trace norm distance:
\be
S_\epsilon(\rho||\sigma) = \inf_{\tau} \left\{S(\rho||\tau):
\tau\ge0, \trace\tau\le1,||\tau-\sigma||_1\le\epsilon\right\}.\label{eq:smoothedRE}
\ee
This form of smoothing has already been applied to Renyi entropies and min- and max-relative entropy \cite{nila,renner,renner12},
giving rise to a quantity with an operational interpretation.
However, the process can equally well be applied to ordinary relative entropy.

By far the most popular modification of the QRE in the quantum case is the \textit{quantum Jensen-Shannon divergence} (QJSD)
\cite{briet09,roga12,fuglede04,lamberti05,roga10},
which has the additional
feature of being symmetric in its arguments. It comes in several flavours: for pairs of states and equal weights,
we have the `vanilla' style:
\bea
\QJS(\rho,\sigma) &=& S(\rho||\half\rho+\half\sigma) + S(\sigma||\half\rho+\half\sigma)\\
&=& S((\rho+\sigma)/2)-(S(\rho)+S(\sigma))/2.
\eea
Here $S(\rho)$ is the von Neumann entropy $S(\rho) = -\trace\rho\log\rho$.
The latter formula allows for a straightforward generalisation to general statistical weights, and to more than two states:
\be
\QJS^{(\pi_1,\ldots,\pi_n)}(\rho_1,\ldots,\rho_n)
=S(\sum_{i=1}^n \pi_i\rho_i) - \sum_{i=1}^n \pi_iS(\rho_i).
\ee
In the context of quantum channels, this quantity is also known as the \textit{Holevo $\chi$} of an ensemble
$\{(\rho_i,\pi_i)\}_{i=1}^n$.

It seems that in the quantum case, Lee's $\alpha$-skew divergence has not been studied before.
%This observation provided the impetus
%behind this paper: to introduce and study the quantum skew divergence.
It was highly rewarding to discover the many interesting properties of the skew divergence,
not to mention the applications presented in this paper.

The skew divergence is closely related to other distinguishability measures.
Firstly, it can be seen as a simplified version of smoothed relative entropy:
to calculate the latter a minimisation problem over states $\tau$ has to be solved.
However, there is a simple canonical choice for $\tau$ that achieves
the same purpose of regularisation but without having to find the exact minimiser.
Namely, we can take that $\tau$ that lies on the $m$-geodesic (mixing geodesic)\footnote{It is not a good idea to choose an $e$-geodesic
(exponential geodesic) here as this once again leads to infinities.}
from $\rho$ to $\sigma$; i.e.\ $\tau=\alpha\rho+(1-\alpha)\sigma$.
In so doing we obtain exactly the skew divergence with $\alpha=\epsilon/||\rho-\sigma||_1$.
For that reason, the skew divergence can be a useful approximation for the smoothed relative entropy. Further study will be devoted to the question
how good this approximation may be.

The skew divergence is also the non-symmetric distinguishability measure underpinning the quantum Jensen-Shannon divergence.
It is therefore not surprising that mathematical results for the skew divergence lead to useful mathematical results for the QJSD
and the Holevo $\chi$. This is the topic of the next and final section.
%%%%%%%%%%%%%%%%%%%%%%%%%%%%%%%%%%%%%%%%%%%%%%%%%%%%%%%%%%%%%%%%%%%%%%%%%%%%%%%%%%%%%%%%%%%%%%%%%%
\section{Inequalities for the Quantum Jensen-Shannon Divergence and Holevo Information\label{sec:holevo}}
The Quantum Jensen-Shannon Divergence (QJS) of $n$ states $\rho_i$, with weights $p_i$, is formally equal to the Holevo information $\chi$,
of the quantum ensemble $\cE=\{(\rho_i,p_i)\}_{i=1}^n$, and is defined as
\be
\QJS^{(p_1,\ldots,p_n)}(\rho_1,\ldots,\rho_n) = \chi(\cE) = S\Big(\sum_i p_i\rho_i\Big) - \sum_i p_i S(\rho_i).
\label{eq:QJSdef}
\ee
We will denote by $\mathbf{p}$ the probability vector $\mathbf{p} = (p_1,\ldots,p_n)$.
Let the averaged state of the ensemble be denoted by $\rho_0 := \sum_i p_i\rho_i$.
It will also be useful to define the \textit{complementary states}
$$
\overline{\rho}_i := \frac{\rho_0-p_i\rho_i}{1-p_i} = \frac{\sum_{j, j\neq i}p_j \rho_j}{1-p_i}.
$$
The Holevo $\chi$ can be rewritten in terms of quantum skew divergences as follows:
\bea
\chi(\cE) &=& \sum_i p_i S(\rho_i || \rho_0) 
= -\sum_i p_i \log (p_i) \SD_{p_i}(\rho_i || \overline{\rho}_i).\label{eq:chiSD}
\eea
From this representation and the bounds on the skew divergence follow several bounds for $\chi$ that improve on existing bounds in the literature.

Let $t_{ij}=T(\rho_i,\rho_j)=||\rho_i - \rho_j||_1 / 2$, the trace distance between signal states $\rho_i$ and $\rho_j$.
Also, let $t=\max_{i,j} t_{ij}$.
From the bound of Theorem \ref{th:SDvT},
$\SD_\alpha(\rho||\sigma)\le T(\rho,\sigma)$, and the convexity of $T$ in each of its arguments, we immediately obtain
\bea
\chi(\cE)
&=& -\sum_i p_i \log (p_i) \SD_{p_i}(\rho_i || \overline{\rho}_i ) \nonumber \\
&\le& -\sum_i p_i \log (p_i) T(\rho_i , \overline{\rho}_i ) \nonumber \\
&\le& -\sum_i p_i \log (p_i) \sum_{j\neq i}\frac{p_j}{1-p_i}  t_{ij} \label{eq:holb1}\\
&\le& H(\mathbf{p}) \;t.\label{eq:holb2}
\eea
In the last line, $H(\mathbf{p}):=-\sum_i p_i \log (p_i)$ is the Shannon entropy of the ensemble's probability vector.
Hence we have shown:
\begin{theorem}
Let $\cE$ be the ensemble $\cE=\{(p_i,\rho_i)\}_{i=1}^n$ with corresponding probability vector $\mathbf{p}=(p_i)_{i=1}^n$.
Let $t$ be the largest of the trace distances $t_{ij}=T(\rho_i,\rho_j)=||\rho_i - \rho_j||_1 / 2$.
Then
\[
\chi(\cE) \le H(\mathbf{p})\; t.
\]
\end{theorem}
This bound combines the well-known bound $\chi(\cE) \le H(\mathbf{p})$ (see, e.g.\ \cite{petzbook}, Th.\ 3.7),
with the bound $\chi(\cE) \le  \log(n) \;t$ of Theorem 14 in \cite{briet09} (only proven there for $n=2$ but clearly true in general), and therefore improves on both.

For binary ensembles, Roga \cite{roga10} proves the following bound on $\chi(\cE)$ in terms of the Uhlmann fidelity $F$ between
the two signal states (see also \cite{roga12} for extensions to more than 2 states):
\be
\chi(\cE) \le S(\sigma),\quad \sigma = \twomat{p}{\sqrt{p(1-p)}F}{\sqrt{p(1-p)}F}{1-p},\label{eq:roga}
\ee
where $F=F(\rho_1,\rho_2) = \trace\sqrt{\sqrt{\rho_1}\rho_2\sqrt{\rho_1}}$.
A numerical investigation showed that  this gives a bound that is sometimes lower in value than (\ref{eq:holb2}), which is in terms
of the trace distance, and sometimes higher.
However, when replacing $t$ by its upper bound $\sqrt{1-F^2}$ in (\ref{eq:holb2}), Roga's bound (\ref{eq:roga}) is always better.
Which bound to choose of course also depends on ease of use and generality.

\bigskip

Now consider two ensembles $\cE$ and $\cE'$ with the same probabilities $p_i$, but different signal states $\rho_i$ and $\rho_i'$,
respectively.
Let $t_i=||\rho_i - \rho_i'||_1 / 2$ be the trace distance between corresponding signal states.
We wish to obtain a bound on $|\chi(\cE) -\chi(\cE')|$ in terms of the $t_i$.
A na{\"\i}ve way to do so would be to use Fannes' continuity bound on
the von Neumann entropy \cite{fannes}. However, this would lead to a bound that is dimension dependent.
Here we show how the two continuity inequalities of the skew divergence (Theorem \ref{th:triangle1})
can be used to obtain a dimension-independent bound.

Define $\rho'_0$, $\overline{\rho}'_i$ analogously as above, $t_0=T(\rho_0,\rho'_0)$ and $\overline{t}_{i}=T(\overline{\rho}_i,\overline{\rho}'_i)$.
The distances $\overline{t}_i$ can be bounded in terms of the $t_j$ as
\be
\overline{t}_i \le \frac{\sum_{j: j\neq i}p_j t_j}{1-p_i} \le \max_{j: j\neq i} t_j\label{eq:tbar}.
\ee
To simplify the formulas, we will express everything in terms of the largest $t_j$, which we denote by $t$.

First consider the difference between terms
\beas
\lefteqn{\SD_{p_i}(\rho_i || \overline{\rho}_i ) - \SD_{p_i}(\rho'_i || \overline{\rho}'_i )} \\
&=&   \SD_{p_i}(\rho_i || \overline{\rho}_i ) - \SD_{p_i}(\rho'_i || \overline{\rho}_i )
    + \SD_{p_i}(\rho'_i || \overline{\rho}_i ) - \SD_{p_i}(\rho'_i || \overline{\rho}'_i ) \\
&\le& \SD_{p_i}(0|1)-\SD_{p_i}(t_i|1)+\SD_{p_i}(t_i|0) + \SD_{p_i}(1|0)-\SD_{p_i}(1|\overline t_i)+\SD_{p_i}(0|\overline t_i) \\
&\le& \SD_{p_i}(0|1)-\SD_{p_i}(t|1)+\SD_{p_i}(t|0) + \SD_{p_i}(1|0)-\SD_{p_i}(1|t)+\SD_{p_i}(0|t) \\
&=& \frac{1}{-\log(p_i)}\left(t\log\frac{p_it+1-p_i}{p_it} + \log\frac{p_i+(1-p_i)t}{p_i}\right).
\eeas
Summing over all terms then yields
\bea
|\chi(\cE) -\chi(\cE')|
&\le& \sum_i p_i t\log\left(1+\frac{1-p_i}{p_i}\;\frac{1}{t}\right) +
\sum_i p_i\log\left(1+\frac{1-p_i}{p_i}t\right). \label{eq:chichi}
\eea
The probabilities $p_i$ can be eliminated by exploiting concavity of the logarithm, giving the promised dimension-independent bound:
\begin{theorem}
Let
$\cE$ and $\cE'$ be two ensembles of $n$ quantum states with the same probabilities $p_i$, but with different states $\rho_i$ and $\rho_i'$,
respectively.
Let $t$ be the largest of $t_{i}:=T(\rho_i,\rho'_i)=||\rho_i-\rho'_i||_1 / 2$.
Then
\bea
|\chi(\cE) -\chi(\cE')|
&\le& t \log(1+(n-1)/t) + \log(1+(n-1)t). \label{eq:chichi2}
\eea
\end{theorem}
For small $t$, this bound is approximated well by $(\log(n-1)+(n-1)-\log t)t$.
%%%%%%%%%%%%%%%%%%%%%%%%%%%%%%%%%%%%%%%%%%%%%%%%%%%%%%%%%%%%%%%%%%%%%%%%%%%%%%%%%%%%%%%%%%%%%%%%%%
\begin{ack}
A substantial part of this work was done at the Institut Mittag-Leffler, Djurs\-holm (Sweden),
during an extended stay at its Fall 2010 Semester on Quantum Information Theory.
I also acknowledge conversations with R.\ Werner, J.\ Oppenheim, B.\ Nachtergaele,
M-B.\ Ruskai and M.\ Shirokov.

Thanks to Tobias Osborne for bringing Bravyi's problem to my attention %(thereby causing no end to headaches)
and to Karel, Micha\"el and Frank for sharing their preprint \cite{karel}.

This work has been supported in part by an Odysseus grant from the Flemish Fund for Scientific Research (FWO).
\end{ack}
%------------------------------------------------------------- BIBLIOGRAPHY

%%%%%%%%%%%%%%%%%%%%%%%%%%%%%%%%%%%%%%%%%%%%%%%%%%%%%%%%%%%%%%%%%%%
\end{document}